\documentclass[comsoc, journal]{IEEEtran}
\usepackage[utf8]{inputenc}
\usepackage[T1]{fontenc}
\usepackage{color, xcolor}
\usepackage{graphicx}
\usepackage{subfig}
\usepackage{float}
\usepackage{enumitem}
\usepackage{hyperref}
\usepackage[noadjust]{cite}
\usepackage{array, makecell, multirow, booktabs}
\newcolumntype{P}[1]{>{\centering\arraybackslash}p{#1}}
\setcellgapes{3.5pt}

\makeatletter
\def\endthebibliography{%
	\def\@noitemerr{\@latex@warning{Empty `thebibliography' environment}}%
	\endlist
}
\newcommand{\rehmat}[1]{\textcolor{black}{#1}}
\newcommand{\rehmatkhan}[1]{\textcolor{black}{#1}}

\begin{document}

\title{\texttt{FedFly}: Towards Migration in Edge-based Distributed Federated Learning}

\author{
	    Rehmat Ullah,
	    Di Wu,
	    Paul Harvey,
	    Peter Kilpatrick,
	    Ivor Spence,
	    and Blesson Varghese
	
	    \thanks{
	    
	   Copyright \copyright~20xx IEEE. Personal use of this material is permitted. Permission from IEEE must be obtained for all other uses, in any current or future media, including reprinting/republishing this material for advertising or promotional purposes, creating new collective works, for resale or redistribution to servers or lists, or reuse of any copyrighted component of this work in other works.
	   
	    This work was supported by funds from Rakuten Mobile, Japan. The last author was also supported by a Royal Society Short Industry Fellowship.
	    
	    R. Ullah, D. Wu and B. Varghese are with the School of Computer Science, University of St Andrews, UK. 
	    
	    P. Kilpatrick and I. Spence are with the School of Electronics, Electrical Engineering and Computer Science, Queen’s University Belfast, UK.
	    
	    P. Harvey is with Autonomous Networking Research \& Innovation Department, Rakuten Mobile, Japan.
	    }	
}


\maketitle

\begin{abstract}
Federated learning (FL) is a privacy-preserving distributed machine learning technique that trains models while keeping all the original data generated on devices locally.
Since devices may be resource constrained, offloading can be used to improve FL performance by transferring computational workload from devices to edge servers. However, due to mobility, devices participating in FL may leave the network during training and need to connect to a different edge server. This is challenging because the offloaded computations from edge server need to be migrated. In line with this assertion, we present \texttt{FedFly}, which is, to the best of our knowledge, the first work to migrate a deep neural network (DNN) when devices move between edge servers during FL training. Our empirical results on the CIFAR-10 dataset, with both balanced and imbalanced data distribution, support our claims that \texttt{FedFly} can reduce training time by up to 33\% when a device moves after 50\% of the training is completed, and by up to 45\% when 90\% of the training is completed when compared to state-of-the-art offloading approach in FL. \texttt{FedFly} has negligible overhead of up to two seconds and does not compromise accuracy. Finally, we highlight a number of open research issues for further investigation. \texttt{FedFly} can be downloaded from \url{https://github.com/qub-blesson/FedFly}.

\end{abstract}


\IEEEpeerreviewmaketitle

\section{Introduction}
\label{sec:introduction}
Internet applications that rely on classic machine learning (ML) techniques gather data from mobile and Internet-of-Things (IoT) devices and process them on servers in cloud data centres. Limited uplink network bandwidth, latency sensitivity of applications and data privacy concerns are key challenges in streaming large volumes of data generated by devices to geographically distant clouds. The concept of Federated Learning (FL) provides privacy by design in an ML technique that collaboratively learns across multiple distributed devices without sending raw data to a central server while processing data locally on devices. 

However, given the limited availability of resources on many devices, performing FL on such devices is impractical due to increased training times~\cite{9084352}. One approach is to leverage the computational resources offered by edge servers (located at the edge of the network) for training. The concept of offloading computations of the ML model that may be a Deep Neural Network (DNN) from a device to an edge server for FL by splitting the ML model has been introduced~\cite{thapa2004splitfed} (this concept is referred to as \textit{edge-based FL}).
However, a major challenge that has not been considered within the context of edge-based FL is \textit{device mobility}. 

Mobile devices participating in edge-based FL may need to move from one edge server to another (for example, a smartphone or a drone moving from the connectivity of one edge node to another). This will in turn affect the performance of edge-based FL and result in large training times~\cite{imteaj2021survey,9415623}. Moving a device without migrating the accompanying training data from an edge server to the destination will result in training for the device having to start all over again on the destination server. This would be inefficient resulting in an increased overall training time~\cite{xia2021survey}. Therefore, there is a need for developing techniques that can move devices while accounting for migrating partially trained FL models of a device from one edge server to another. 

Research on device mobility has been considered in the context of migration. Migration on the edge has been investigated in the literature, more specifically by exploring VM migration~\cite{zhang2018survey} and container migration~\cite{singh2021taxonomy, 7980161}. However, migration in edge-based FL is minimally considered. This paper presents \texttt{FedFly} that addresses the mobility challenge of devices in edge-based FL, and the \textbf{\textit{key research contributions}} are:

(1) The \textit{technique for migrating DNNs in edge-based FL}, which to the best of our knowledge is the first time to be considered in the context of edge-based FL. When a device moves from an edge server to a destination server after 50\% of FL training is completed, then the training time using the \texttt{FedFly} migration technique is reduced by up to 33\% compared to the training time when restarting training on the destination server. Similarly, 45\% reduction is obtained when a device moves to a destination server after 90\% FL training is completed. It is noted that the original accuracy is maintained. 

\rehmat{(2) The \textit{implementation and evaluation of \texttt{FedFly} in a hierarchical cloud-edge-device architecture} that validates the migration technique of edge-based FL on a lab-based testbed. The experimental results are obtained from a lab-based testbed that includes four IoT devices, two edge servers, and one central server (cloud-like) running the VGG-5 DNN model. The evaluation is done on both a balanced (equal data distribution) and an imbalanced dataset (unequal data distribution). The empirical findings show that \texttt{FedFly} has a negligible overhead of up to two seconds on the testbed. It is further noted that the accuracy is preserved even when data on devices is imbalanced and the most significant node(s) (i.e., nodes with majority of data) move across edge servers.}

The rest of this paper is organized as follows: Section~\ref{sec:background} introduces the concepts of FL and offloading in FL. Section~\ref{sec:problem} presents  the motivation for \texttt{FedFly}. Section~\ref{sec:proposed} proposes the migration technique for edge-based FL. Section~\ref{sec:evaluation} presents the performance analysis of \texttt{FedFly}. Section~\ref{sec:conclusion} concludes the paper and highlights directions for future research.

\section{Background}
\label{sec:background}

This section provides an overview of FL and highlights the benefits of offloading ML computations on to edge servers. 

FL~\cite{mcmahan2017communication} is a privacy-preserving technique in which an ML model is collaboratively trained across several participating distributed devices. All data generated by a device that is used for training resides on local devices. In an FL system, the server initiates a global model and distributes the model parameters to all connected devices. Then each device trains a local version of the ML model using local data. Instead of sending the raw data to the server, the local model parameter updates are sent up to the server. Subsequently, the server computes a weighted average using the parameter updates on the server using the Federated Averaging (FedAvg) algorithm \cite{mcmahan2017communication} to obtain a new set of parameters for the global model. The updated global model is then sent back down to each device for the next round of training by the edge server. The entire process is repeated until the model converges~\cite{gao2020end}.

In practice, running FL across resource constrained devices, for example in an IoT environment, will result in large training times. Therefore, the concept of partitioning and offloading the ML model, for example for a DNN has been explored for performance efficiency~\cite{9302776}. Split Learning (SL)~\cite{vepakomma2018split} is one ML technique that leverages this concept. 

\rehmat{In SL, a DNN is partitioned across the device and server. The DNN layer after which the model is partitioned is referred to as the split layer. The device trains the model up to the split layer and then sends the split layer activation (referred to as smashed data) to the server. The server trains the remaining layers of the DNN using the smashed data. The server performs back-propagation up to the split layer and sends the gradients of the smashed data to the devices. The devices use the gradients to perform back-propagation on the rest of the DNN.}

However, when multiple devices participate in SL, the devices are trained in a sequential round robin fashion whereby only one device is connected to the server at a time.   This limitation is overcome by SplitFed~\cite{thapa2004splitfed} and FedAdapt~\cite{wu2021fedadapt}. SplitFed and FedAdapt allow for simultaneous training of all participating devices and at the same time leverage on partitioning the DNN to alleviate the computational burden of training on the device. In addition to the underlying approaches of SplitFed, FedAdapt incorporates a reinforcement learning approach to dynamically identify the DNN layers that need to be offloaded from the device to the edge based on the operational conditions of the environment. In this paper, SplitFed is considered as the baseline. 


SplitFed reduces the amount of computation carried out on the device and is faster than classic SL. However, it is limited in that the challenge of device mobility during training has not been considered. Currently, there is no research in the literature that considers the migration of edge-based FL when devices move between edge servers. The next section highlights the key challenges when using SplitFed. 

\begin{figure*}[t]
	\centering
	\includegraphics[width=0.85\linewidth]{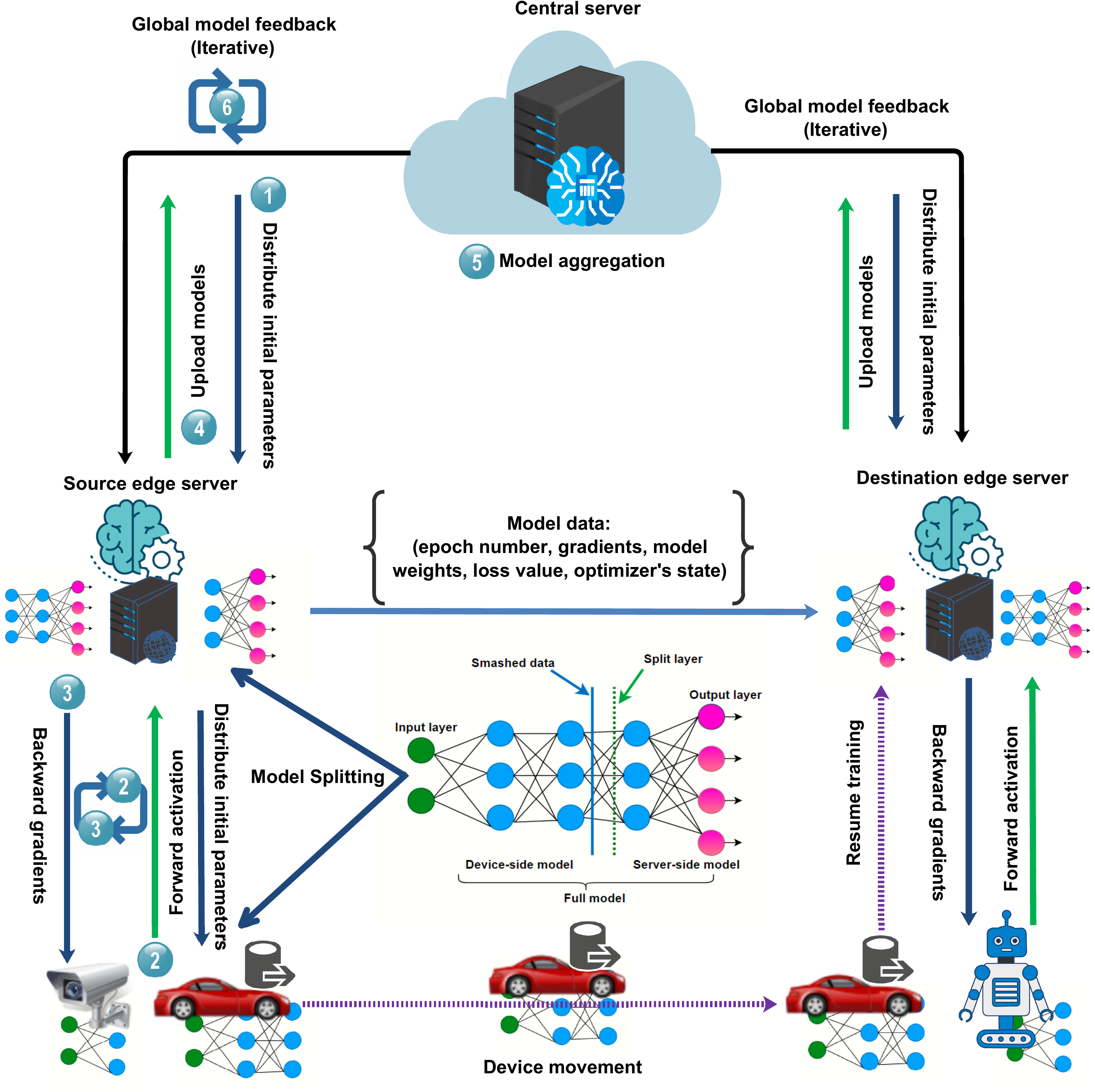}
	\caption{System of \texttt{FedFly}.}
	\label{fig:FedFly}
\end{figure*}

\section{Impact of Device Mobility on Edge-based Federated Learning}
\label{sec:problem}

This section considers the impact of \textit{device mobility} on the training time in edge-based FL. 
Three contributing factors, namely model training, imbalanced data distribution and frequency of device mobility are considered. 

\textit{Model training:} Due to mobility, a device participating in FL may disconnect from one edge server and will need to connect to another server at any stage during training. \rehmat{For example, in the early stages of training, if a device moves, restarting training on a different edge server may result in a small increase in training time. However, if the device had completed a larger portion of its training on an edge server before the device moved, then the training time would significantly increase.} A migration mechanism is required so that mobile devices can resume training on the destination edge server rather than starting over.

\textit{Imbalanced data distribution}:
In a real edge-based FL system, some devices may have more data than others due to frequent use of specific services or have more resources such as memory~\cite{imteaj2021survey, mcmahan2017communication}. Consequently, these  devices will make a significant contribution to the quality (overall accuracy) of the global model. However, devices that generate a large amount of data cannot be removed from contributing to training since the eventual accuracy of the global model will be adversely affected.  Furthermore, devices with more data will require more training time. As a result, restarting training for the device after it has moved to a different edge server will increase the training time. A migration mechanism that allows such devices to resume training (rather than restarting from the beginning) when moving between edge servers is required to reduce training time while not compromising the global model accuracy.

\textit{Frequency of device mobility}:
The frequency with which devices may move between edge servers can have an impact on  training time. If the devices move frequently during training, the overall training time will increase because  training will need to be restarted on each device after it has moved to a different edge server. 

In this paper, we present \texttt{FedFly}, that aims to address the \textit{device mobility} challenge by taking into account the above factors for reducing the training time and maintaining the accuracy of the global model as close to that in classic FL.

	\begin{figure*}[t]
	\centering
	\includegraphics[width=0.85\linewidth]{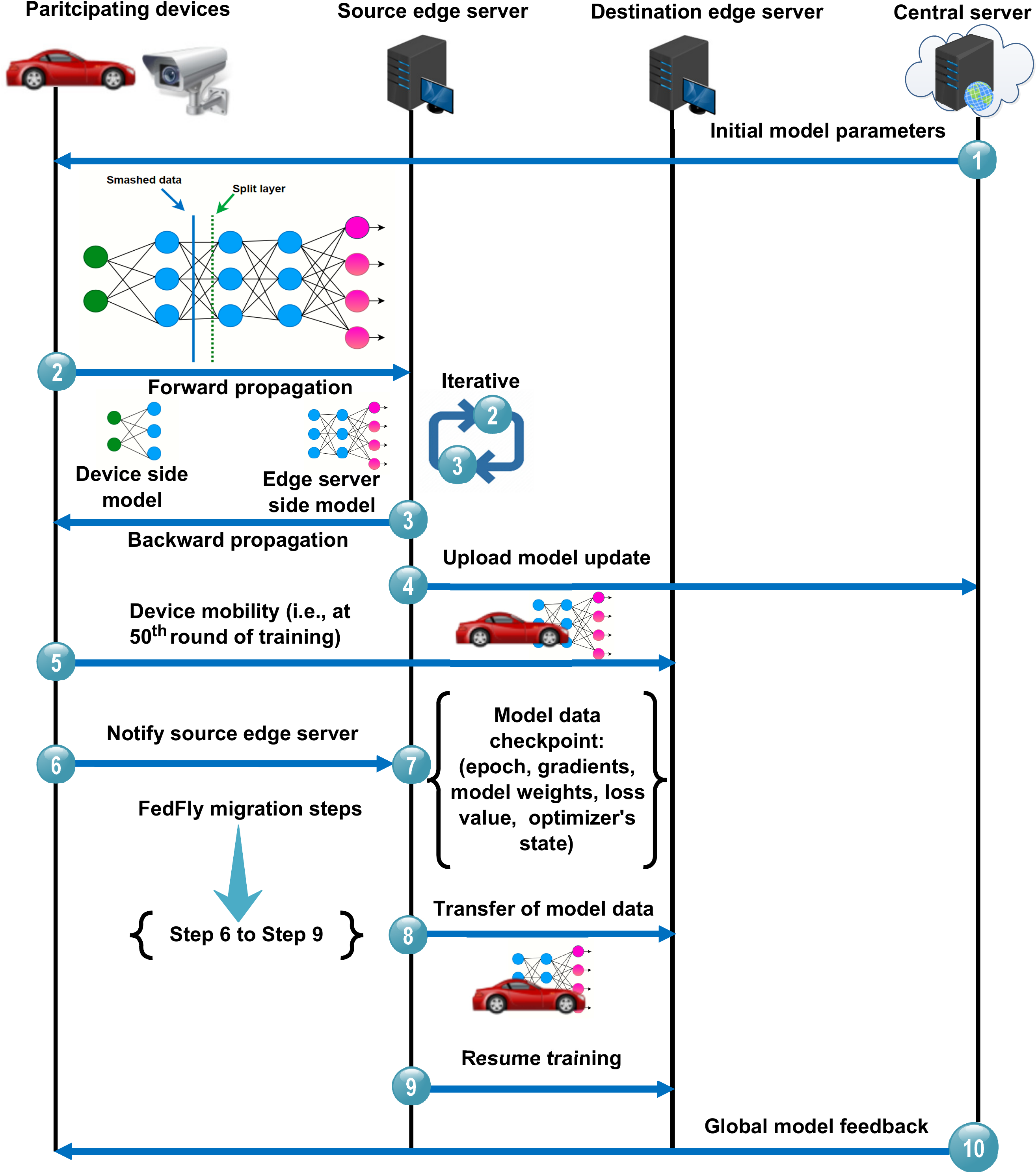}
	\caption{Sequence diagram of \texttt{FedFly}.}
	\label{fig:SequenceDiagram}
\end{figure*}

\section{\texttt{FedFly} for Migration in Hierarchical Edge-based Federated Learning}
\label{sec:proposed}
This section presents \texttt{FedFly} (\url{https://github.com/qub-blesson/FedFly}), the edge-based distributed FL system that caters for mobility of devices. A hierarchical structure that comprises three entities, namely devices, edge servers, and a central server (cloud-like) is considered. The \texttt{FedFly} system is shown in Figure~\ref{fig:FedFly}. The following highlights the steps in relation to distributed FL and the mobility of devices within the \texttt{FedFly} system: 

\textit{Central server initialization:} When training begins, the central server initializes the global model parameters and distributes them to the edge servers. The model parameters are received by the edge servers and passed to the participating devices (Step 1). The training  on the devices begins when the devices receive the model parameters from the servers.

\textit{Splitting Deep Neural Networks:} When the model is initialized, the DNN that would in classic FL run on a device is split between device and the edge server. 
After all devices and edge servers complete local training on the data generated by the device, i.e., forward and backward propagation (Step 2 and Step 3), the local model updates are sent to the central server for global model aggregation (Step 4). A complete forward and backward propagation corresponds to one local epoch (an epoch refers to one complete cycle of an entire dataset on a device through the neural network) of a device for all local data of that device. The central server aggregates the model (Step 5), and then the updated parameters of the global model are sent back to the edge servers and devices for training for a next round of FL training (Step 6). 
		
At any point during training, it is possible for a device to move between edge servers. Figure~\ref{fig:SequenceDiagram} shows the sequence of activities initiated by \texttt{FedFly} when a device needs to move from source edge server to the destination edge server. Assume that a device disconnects from the source edge server after the $50^{th}$ round of training. When a device connects to the destination server without using a migration mechanism, all the training is lost until the $50^{th}$ round, and training is restarted on the destination edge server. This is because the destination edge server does not have a copy of the model that was trained on the source edge server. It is necessary to migrate the model data from the source edge server to the destination edge server before training can resume. 

\texttt{FedFly} overcomes the mobility challenge by migrating model data from the source edge server to the destination edge server. There are three steps that are considered in \texttt{FedFly} when a device starts moving during FL training. 

\textit{Notify edge server:} When a device starts to move, it notifies the source edge server to prepare data that needs to be migrated to the destination edge server (Step 6). \rehmat{In this article it is assumed that the moving device knows when to disconnect from the source edge server.} 

\textit{Model data checkpoint:} The source edge server creates a data checkpoint that includes the epoch number, gradients, model weights, loss value, and state of optimizer (such as Gradient Descent) (Step 7). \rehmat{The checkpointed data is transferred via a socket to the destination edge server (Step 8).}

\rehmat{\textit{Resume training:} At the destination edge server, the checkpointed data is received via a socket. When a device connects to the destination edge server, training is resumed from the point where the device started moving at the source edge server (Step 9). } 
 
There are several possible ways to transfer model data between edge-servers. In \texttt{FedFly}, the source edge server transfers data directly to the destination edge server, after which the device resumes training. However, in practice the two-edge servers may not be connected or may not have the permission to share data with each other. In this case, the device can then transfer the checkpointed data between edge servers.

\section{Evaluation}
\label{sec:evaluation}
This section first describes the experimental setup, including the lab-based testbed used for carrying out experiments, and then substantiates the key claims of \texttt{FedFly} by presenting and analysing the results obtained.
\subsection{Experimental Setup}

The testbed includes four devices, two edge servers and one central server. The devices are: (i) two Raspberry Pi 4 (Pi4\_1, and Pi4\_2) Model B with 1.5GHz quad-core ARM Cortex-A72 CPU, 4GB RAM and 32GB storage, and (ii) two Raspberry Pi 3 (Pi3\_1, and Pi3\_2) Model B with 1.2GHz quad-core ARM Cortex-A53 CPU, 1GB RAM and 32GB storage. The edge servers comprise: (i) a 2.3GHz quad-core Intel i5 CPU, 8GB RAM and 256GB storage, and (ii) a 2.3GHz quad-core Intel i7 CPU, 16GB RAM and 500GB storage. The central server has a 2.9GHz quad-core Intel i5 CPU, 16GB RAM and 1TB storage. All Raspberry Pis have the same version of Raspbian GNU/Linux 10 (Buster) operating system, Python version 3.7 and PyTorch version 1.4.0. The edge servers and the central server have the same version of Python and PyTorch using Anaconda. \rehmatkhan{All devices are connected to the servers in a Wi-Fi network with an average available bandwidth of 75Mbps}.

The DNN model used is VGG-5~\cite{simonyan2014very} and the CIFAR-10~\cite{krizhevsky2009learning} dataset is used as input with size $3@32\times32$ and a batch size of 100 is used for all experiments. The CIFAR-10 dataset contains 50K training and 10K testing samples that consist of color images of ten objects (classes), including plane, car, bird, cat, deer, dog, frog, horse, ship, and truck. The standard FedAvg~\cite{mcmahan2017communication} aggregation method is used, and the model parameters are updated using Stochastic Gradient Descent (SGD), with a learning rate of 0.01 and a momentum of 0.9.

\subsection{Empirical Results and Discussion}
In this section, we demonstrate the performance of \texttt{FedFly} by comparing it with SplitFed in terms of device training time and model accuracy. We validate our claims using balanced and imbalanced datasets at various stages (i.e., 50\% and 90\%) of FL training.

\begin{figure}[]
	\centering
	\subfloat[]{\includegraphics[width=0.47\textwidth]{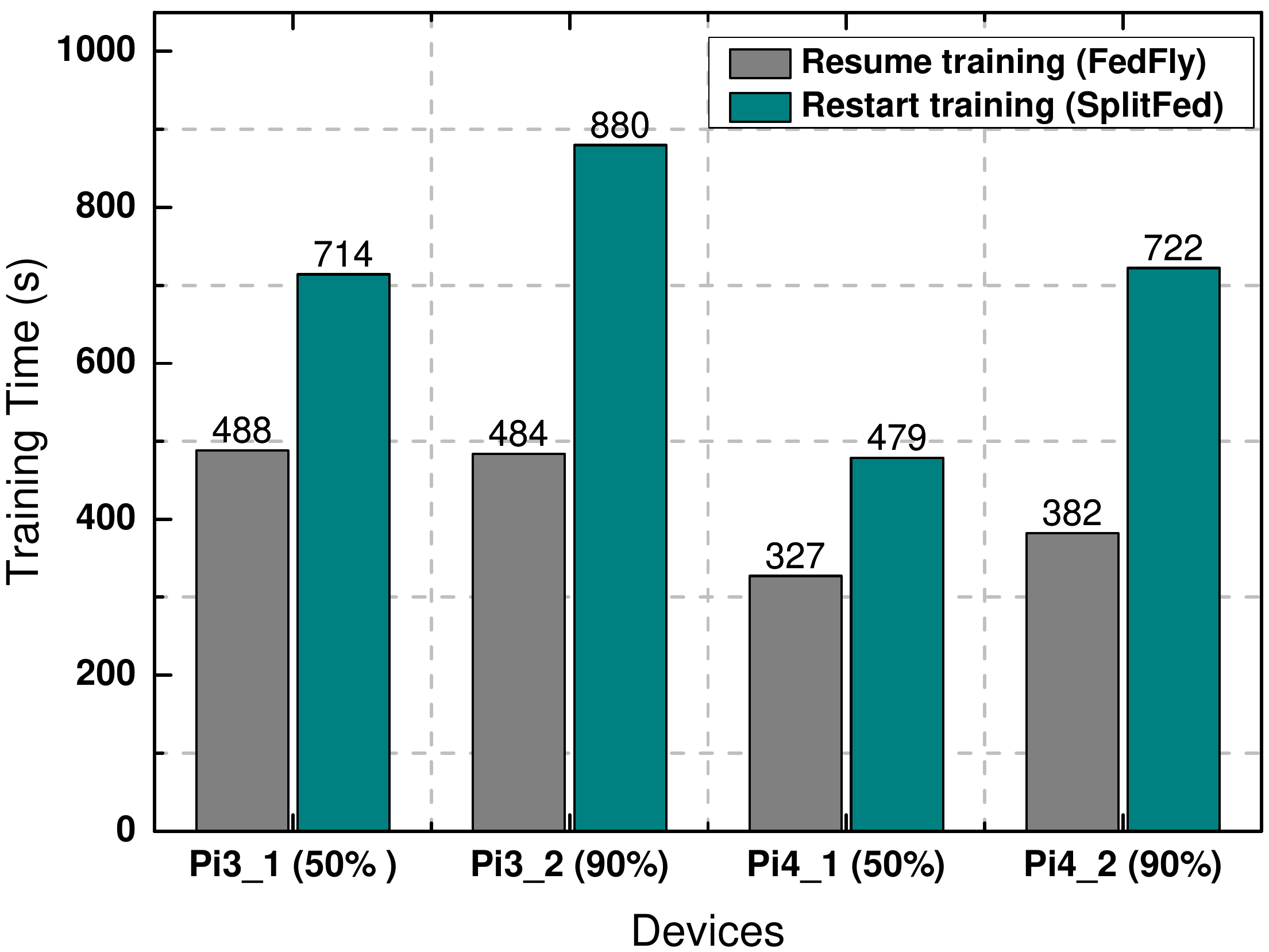} }
	\\             
	\subfloat[]{\includegraphics[width=0.47\textwidth]{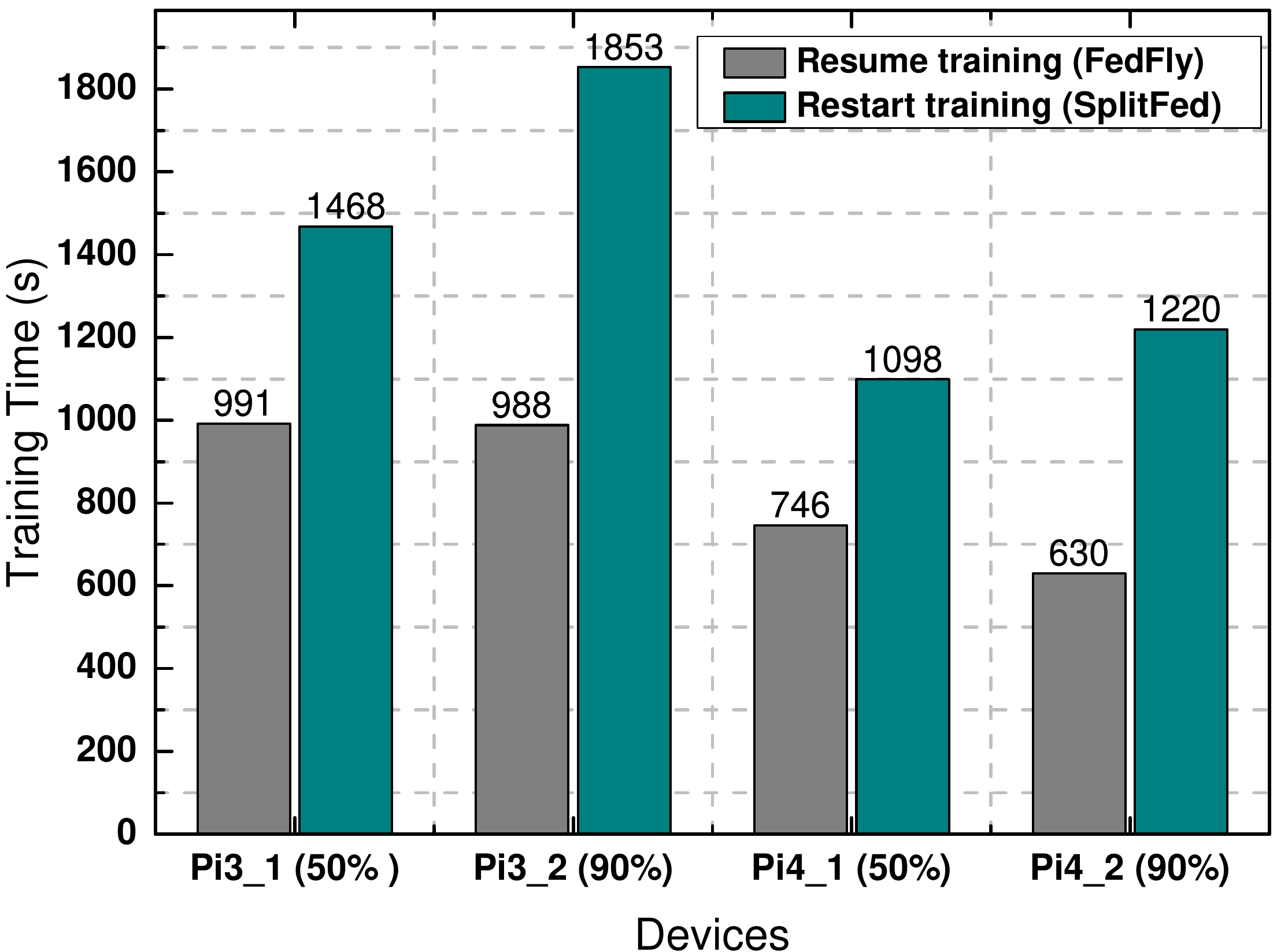} }
	\\
\subfloat[]{\includegraphics[width=0.47\textwidth]{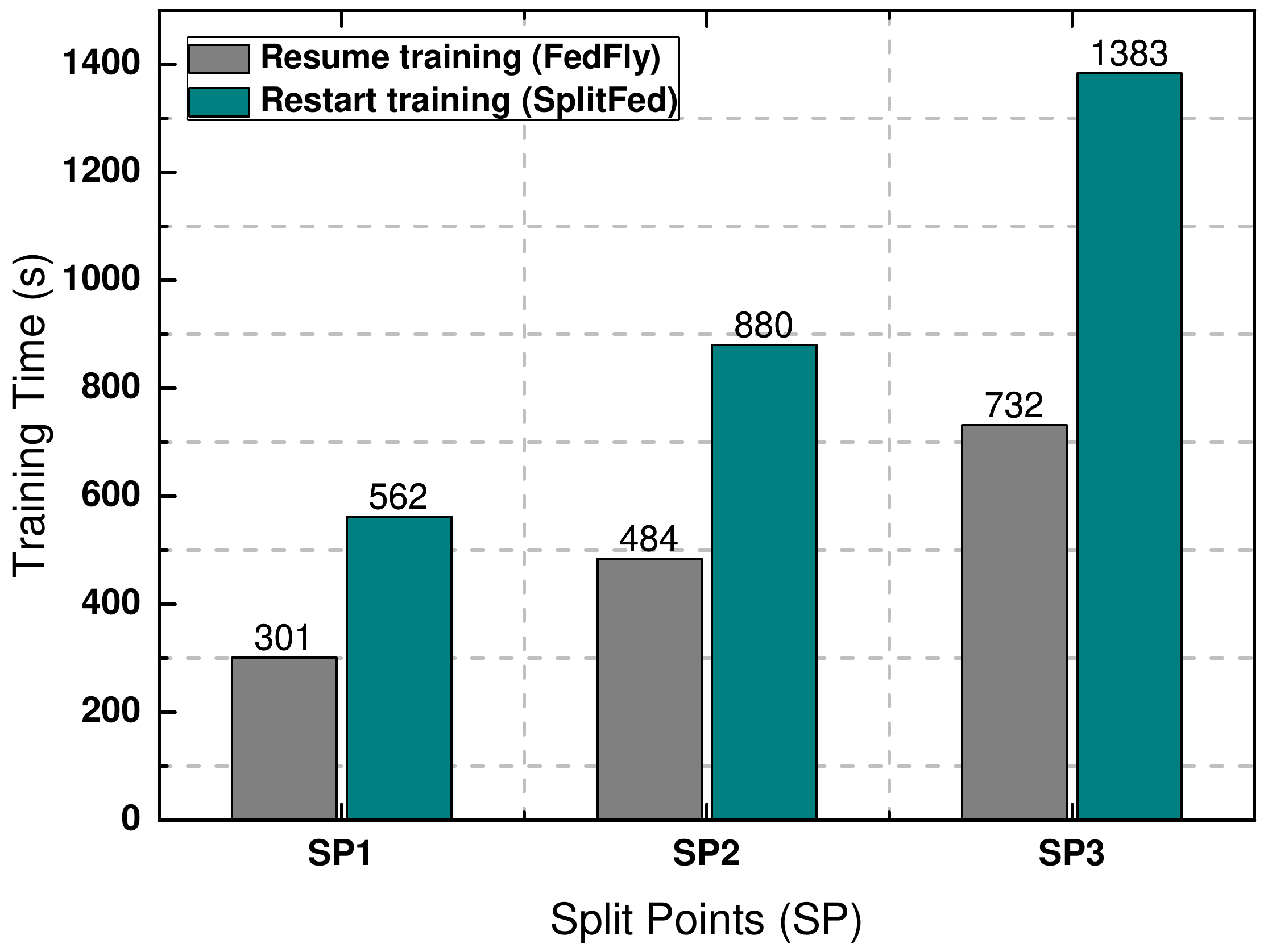} }
	\caption{(a) Device training time per round when 25\% of the dataset is required for training on a mobile device, (b) Device training time per round when 50\% of the dataset is required for training on a mobile device
\rehmatkhan{(c) Device training time per round by varying SPs with 25\% of the dataset on a mobile device and at 90\% of the FL training.}}
	\label{fig:trainingtime}
\end{figure}

\textbf{Effect of mobility on device training time:}
When a device moves between edge servers, factors such as training stage and the dataset available on the device can affect training time. In this experiment, we validate the training time claim by generating 25\% and 50\% of the data required for training on a single device (i.e., Pi3\_1, Pi3\_2, Pi4\_1 and Pi4\_2) with training stages at 50\% and 90\% as shown in Figure \ref{fig:trainingtime} (a) and Figure \ref{fig:trainingtime} (b).

Figure \ref{fig:trainingtime} (a) shows the effects of device mobility on device training time when 25\% of the dataset is required for training on a single device, as well as device movement when 50\% and 90\% of the training is completed. It is evident from Figure \ref{fig:trainingtime} (a) that \texttt{FedFly} always outperforms SplitFed, in which the training is restarted at the destination edge server. When we move Pi3\_1 when 50\% of the training is done, the training time is reduced by up to 33\% per round. However, when we move Pi3\_2 with the same dataset but 90\% of the training is completed, the training time is reduced by up to 45\% per round. We also move devices (Pi4\_1 and Pi4\_2) when 50\% and 90\% of the training is done, and the training time is reduced by up to 33\% and 45\% per round, respectively. 

Figure~\ref{fig:trainingtime} (b), shows the effects of device mobility on device training time when 50\% of the dataset is required for training on a single device, as well as device movement when 50\% and 90\% of the training is completed. It can be seen in Figure~\ref{fig:trainingtime} (b) that training time on devices is longer than on devices in Figure~\ref{fig:trainingtime} (a). This is due to the fact that 50\% of the dataset is used for training on mobile devices, which is comparably larger than used for devices in Figure~\ref{fig:trainingtime} (a). It has been demonstrated from Figure~\ref{fig:trainingtime} (a) and Figure~\ref{fig:trainingtime} (b) that \texttt{FedFly} can save a significant amount of training time when compared to SplitFed. 

\rehmatkhan{Figure~\ref{fig:trainingtime} (c) highlights the system performance with device mobility by varying the split points (SP). SP1 denotes the first convolutional layer on devices, SP2 denotes the first two convolutional layers on devices, and SP3 denotes the first three convolutional layers on devices, with the remaining layers on edge-servers. It should be noted that in the experiments illustrated in Figure \ref{fig:trainingtime} (a) and Figure \ref{fig:trainingtime} (b), all devices and edge servers have fixed split points (i.e., SP2). Figure~\ref{fig:trainingtime} (c) depicts that SPs impact the system performance, in terms of training time. By changing the SPs from SP1 to SP3, we note a significant increase in training time. This is because as the number of layers (i.e., computation) on devices and servers increases or decreases, the training time on devices or servers increases or decreases accordingly. 
In all cases, \texttt{FedFly} saves a significant amount of training time when compared to SplitFed. The transfer time is still up to two seconds. This is because the VGG-5 model is used in the experiments, and the data that is checkpointed did not change significantly by varying SPs.}


\textbf{Effect of mobility on global accuracy:}
In this experiment, we verify the accuracy of the global model when a device moves frequently between edge servers. 

\begin{figure}[H]
	\centering
	\includegraphics[width=0.47\textwidth]{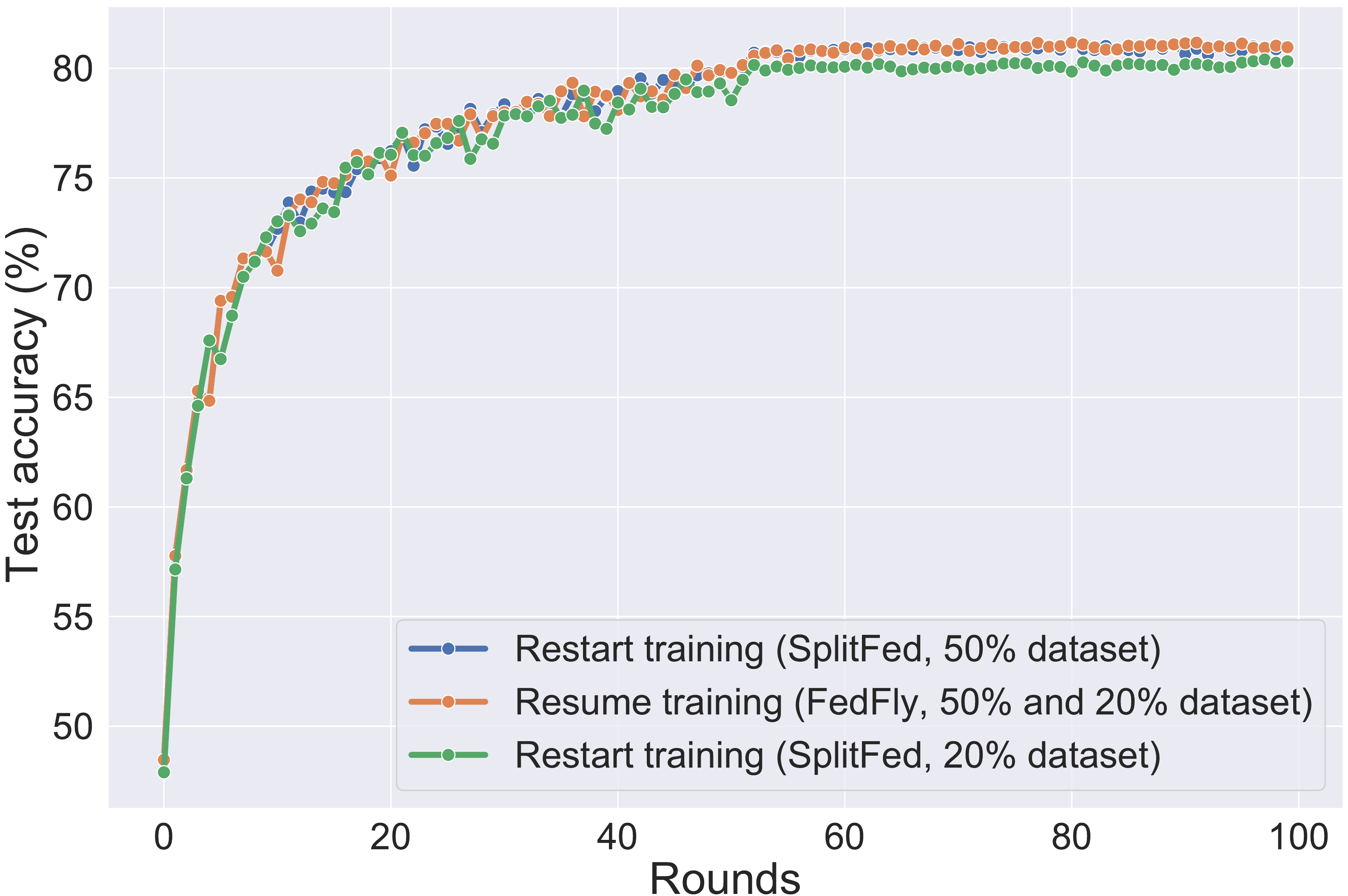}
	\caption{Global accuracy when 20\% and 50\% of datasets are required for training on a mobile device for 100 rounds of training.}
	\label{fig:GlobalAccuracy}
\end{figure}

We ran this experiment for a total of 100 rounds, with a mobile device holding 20\% of the dataset and 50\% of the dataset. We move the device at various rounds during 100 rounds of training, such as at the $10^{th}$, $20^{th}$, $30^{th}$, $40^{th}$, $50^{th}$, $60^{th}$, $70^{th}$, $80^{th}$, and $90^{th}$ rounds.
Figure \ref{fig:GlobalAccuracy} clearly shows that there is no effect on accuracy. \texttt{FedFly} and SplitFed both maintain accuracy when a device moves between edge servers holding 20\% and 50\% of the datasets. In the case of SplitFed, the training is restarted at the destination edge server without any accuracy loss. This is because the device obtains the updated model parameters from the central server and restarts training at the destination edge server. For example, if a device moves at the $10^{th}$ round, the central server has the updated model parameters until the $10^{th}$ round, and when a device connects to the destination edge server, it receives updated parameters from the central server. Only the training is restarted, which increases the training time but has no effect on accuracy. 

\texttt{FedFly}, on the other hand, transfers the data to the destination edge server, where training is resumed and maintains the same level of accuracy as SplitFed. The training, however, is not repeated at the destination edge server.

\subsection{Summary of the evaluation results}

\texttt{FedFly} performance is affected by a number of factors, including i) balanced and imbalanced datasets on devices, \rehmatkhan{ii) varying the SPs,} iii) the frequency with which devices move; and iv) the model training stages. Our experimental results provide the following insights:
\begin{itemize}
    
    \item In comparison to SplitFed, \texttt{FedFly} reduces the training time per round by up to 33\% when a device moves after 50\% of the training is completed, and by up to 45\% when 90\% of the training is completed.

    \item \texttt{FedFly} maintains global accuracy as does SplitFed and there is no accuracy loss. 
    
    \item \texttt{FedFly} results in up to two seconds overhead, which is the time it takes to transfer data between edge servers during migration. This overhead is negligible when compared to the device training time when training is restarted at the destination server. \rehmat{The reduction in training time and overhead reported in this paper are based on experiments carried out on the lab-based testbed.}
\end{itemize}

\section{Conclusion and Future Research Directions}
\label{sec:conclusion}
The FL system is hindered by two major issues: training time and accuracy. This becomes more challenging when a device moves during FL training and especially when a DNN is partitioned between device and edge server. This paper has proposed \texttt{FedFly}, which for the first time addresses the \textit{device mobility} challenge during FL training, particularly in edge-based FL. We develop a prototype on a lab-based testbed, that upholds and validates our claims in terms of training time and accuracy using balanced and imbalanced datasets when compared to state-of-the-art SL approach called SplitFed. Our empirical results reveal that \texttt{FedFly} introduces a negligible overhead but saves a significant amount of training time while maintaining accuracy. 

\subsubsection*{\textbf{Future Research Directions}}
We develop \texttt{FedFly} for migration in edge-based distributed FL \textemdash but this is only the tip of the iceberg of the opportunities it makes available. What follows are a few research questions that we may further investigate.

\rehmat{\textit{Multiple devices mobility:} Further challenges may occur in the FL setting if multiple devices try to move at the same time with various data distribution at each node. The impact of a large number of devices on training time and accuracy will be investigated further in order to realise migration in practical FL systems.}

\textit{Hardware heterogeneity:} In \texttt{FedFly}, we perform migration in a homogeneous environment, i.e., the hardware at the edge servers is of the same instruction set architecture (ISA). However, in practical scenarios, edge servers are often built with CPUs of different ISAs. As a result, a DNN model that has been natively trained for one ISA cannot be moved to another, making migration to the destination edge server difficult. Migration at runtime across edge servers featuring CPUs of different ISAs, such as ARM and x86, requires further investigation. 

\textit{Neural network optimization:} In practice, the destination edge server may not have enough resources to run the DNN model, meaning that the destination edge server resource is not equivalent to the source edge server resource. How to move DNN on the fly so that the DNN model can run on the destination edge server with limited resources and how to optimise the DNN without impacting its accuracy may be further investigated.

\textit{Asynchronous training:} \texttt{FedFly} currently focuses on synchronous training in edge-based distributed FL. However, the practical FL scenario shows significant heterogeneity in terms of computation resources, hardware, dataset distribution, and communication, etc. It would be worthwhile to investigate the migration issues for asynchronous training in edge-based distributed FL.

\rehmat{\textit{Communication overhead:}  \texttt{FedFly} does not impose any communication challenges, as training from the source edge server is resumed with a 2 second overhead at the destination edge server. However, communication challenges may arise as a result of the hierarchical cloud-edge-device architecture in which \texttt{FedFly} operates since the volume of communication between the cloud, edge servers and devices increase. This may result in a higher communication overhead since model parameters are frequently shared between the cloud to edge to device and vice-versa. Efficient mechanisms for reducing communication overhead between devices, edge servers, and the cloud will be considered in the future.}

\bibliographystyle{IEEEtran}
\bibliography{Ref}

\vskip -2.5\baselineskip plus -1fil

\begin{IEEEbiographynophoto}{Rehmat Ullah} 
is a research fellow  at the University of St Andrews, UK. 
His research focuses on the broader areas of network and distributed systems, particularly edge computing and information centric networking, with a recent focus on federated learning for edge computing systems.  More information is available from \url{www.rehmatkhan.com}.

\end{IEEEbiographynophoto}
\vskip -2.5\baselineskip plus -1fil
\vspace{0.2cm}

\begin{IEEEbiographynophoto}{Di Wu} is currently pursuing a PhD degree in computer science at University of St Andrews, UK. 
His major interests are in the areas of federated learning, distributed machine learning, edge computing, model compression, and Internet-of-Things.

\end{IEEEbiographynophoto}
\vskip -2.5\baselineskip plus -1fil
\vspace{0.2cm}

\begin{IEEEbiographynophoto}{Paul Harvey} is one of
the original founders of the Autonomous Networks
Research and Innovation Lab in Rakuten Mobile,
Japan, and is a co-chair in the ITU focus group
on autonomous networks. He is Research Lead at the Autonomous Networking Research and Innovation Department, Rakuten Mobile, Japan.  

\end{IEEEbiographynophoto}
\vskip -2.5\baselineskip plus -1fil
\vspace{0.2cm}

\begin{IEEEbiographynophoto}{Peter Kilpatrick} is a Reader in computer science at Queen's University Belfast, UK. His interests include parallel programming models and cloud and edge computing.
\end{IEEEbiographynophoto}

\vskip -2.5\baselineskip plus -1fil
\vspace{0.2cm}

\begin{IEEEbiographynophoto}{Ivor Spence} is a Reader in computer science at Queen's University Belfast, UK, where he leads the artificial intelligence (AI) research theme with a focus on heterogeneous computing systems for AI.

\end{IEEEbiographynophoto}
\vskip -2.5\baselineskip plus -1fil
\vspace{0.2cm}

\begin{IEEEbiographynophoto}{Blesson Varghese} 
is  a Reader in computer science at the University of St Andrews, UK, and the Principal Investigator of the Edge Computing Hub. 
His recent interests are at the intersection of the cloud-edge-device continuum and machine learning. More information is available from \url{www.blessonv.com}.

\end{IEEEbiographynophoto}

\end{document}